\documentclass[aps,showpacs,PRB,twocolumn,superscriptaddress]{revtex4-2}
\makeatletter

\newcommand{\Rmnum}[1]{\expandafter\@slowromancap\romannumeral #1@}
\makeatother
\usepackage[english]{babel}
\usepackage{amsmath}
\usepackage{graphicx}
\usepackage{subfigure}
\usepackage{color}
\usepackage{amsfonts}
\usepackage{tikz}
\usepackage{esint}
\usepackage{mathrsfs}
\usepackage{verbatim}

\newcommand{\bea}{\begin{eqnarray}}
\newcommand{\eea}{\end{eqnarray}}
\newcommand{\down}{\downarrow}
\newcommand{\up}{\uparrow}

\begin{document}

\title{Accessing Excitation of Many-body Systems via Single-Mode Approximation within Quantum Monte Carlo Simulations}

\author{Yan Liu}
\affiliation{Department of Physics and State Key Laboratory of Surface Physics, Fudan University, Shanghai 200438, China}

\author{Kemeng Wu}
\affiliation{Department of Physics and State Key Laboratory of Surface Physics, Fudan University, Shanghai 200438, China}

\author{Shutao Liu}
\affiliation{Department of Physics and State Key Laboratory of Surface Physics, Fudan University, Shanghai 200438, China}

\author{Yan-Cheng Wang}
\affiliation{Collaborative Center for Physics and Chemistry, Institute of International Innovation, Beihang University, Hangzhou 311115, China}
\affiliation{Tianmushan Laboratory, Hangzhou 311115, China}

\author{Jie Lou}
\email{louejie@fudan.edu.cn}
\affiliation{Department of Physics and State Key Laboratory of Surface Physics, Fudan University, Shanghai 200438, China}
\affiliation{Collaborative Innovation Center of Advanced Microstructures, Nanjing 210093, China}

\author{Zheng Yan}
\email{zhengyan@westlake.edu.cn}
\affiliation{Department of Physics, School of Science and Research Center for Industries of the Future, Westlake University, Hangzhou 310030,  China}
\affiliation{Institute of Natural Sciences, Westlake Institute for Advanced Study, Hangzhou 310024, China}
\affiliation{Lanzhou Center for Theoretical Physics $\&$ Key Laboratory of Theoretical Physics of Gansu Province, Lanzhou University, Lanzhou, Gansu 730000, China}

\author{Yan Chen}
\email{yanchen99@fudan.edu.cn}
\affiliation{Department of Physics and State Key Laboratory of Surface Physics, Fudan University, Shanghai 200438, China}
\affiliation{Collaborative Innovation Center of Advanced Microstructures, Nanjing 210093, China}

\begin{abstract}
We extend the single-mode Approximation (SMA) into quantum Monte Carlo simulations to provides an efficient and fast method to obtain the dynamical dispersion of quantum many-body systems. Based on stochastic series expansion (SSE) and its projector algorithms, the SMA + SSE method can simply extract the dispersion of the dynamical dispersion in the long wave-length limit and the upper bound of the dispersion elsewhere, without external calculations and high technique barriers. Meanwhile, numerical analytic continuation methods require the fine data of imaginary time correlations and complex programming. Therefore, our method can approach the excitation dispersion of large systems, e.g., we take the two-dimensional Heisenberg model on a $512 \times 512$ square lattice. We demonstrate the effectiveness and efficiency of our method with high precision via additional examples. We also demonstrate that SMA combined with SSE goes beyond spin-wave theory with numerical results. We further illustrate that SMA is able to extract useful information in strongly correlated systems with competing states.
\end{abstract}

\newcommand{\myref}[1]{Eq.(\ref{#1})}
\maketitle
\section{introduction}
Strongly correlated systems emerge with many novel phenomena and thus attract much attention. Usually, exotic quantum states with peculiar behaviors do not thoroughly exhibit themselves in small systems due to finite-size effects. The exponentially increasing degree of freedom of the Hilbert space hinders further understanding of quantum many-body systems. This stimulates people to derive new approaches to more extensive system sizes. Quantum Monte Carlo (QMC) is a powerful numerical tool for dealing with complex systems, especially with a high degree of freedom~\cite{Gubernatis2016QMCmethods,Sandvik2019SSE}.

Generally, there are two main branches of QMC methods. The first branch uses stochastic processes to simulate the finite temperature partition function of quantum many-body systems. This branch includes algorithms like stochastic series expansion (SSE)~\cite{Sandvik1991SSE,Sandvik1999SSE,Syljuasen2002DirectedLoop,ZY2019Dimer,ZY2020DimerImproved,Sandvik2019SSE,Nisheeta2021SSE} and path integral~\cite{Prokof1998Worm,Boninsegni2006Worm,Boninsegni2006Worm_2,Krzakala2008Path}. The other one performs the ground state wave function at zero temperature, such as diffusion Monte Carlo~\cite{Kosztin1996Diffusion,Syljuaasen2005Diffusion,Syljuaasen2005RandomWalk,Syljuaasen2005Diffusion_2,Syljuaasen2006Diffusion} and Green's function Monte Carlo~\cite{Trivedi1990Green,Trivedi1989Green,Arnow1982Green,Lee1992Green}.

Although knowledge of the ground state is always what people seek in the first place, excited states and energy spectrum, which carry information on the energy gap and dynamical exponent $z$, also play a crucial role in our understanding of the system. Experiments like neutron scattering have been performed to explore the excitations in antiferromagnetic materials~\cite{dalla2015excitation,sala2021excitation}. Obtaining the excitation information of many-body systems is one of the most challenging tasks in QMC simulations. Some numerical analytical continuation (NAC) methods like maximum entropy method and stochastic analytic continuation (SAC)~\cite{Gull1984MaxEntropy,Sandvik1998AC,Beach2004MaxEntropy,Syljuaasen2008AC,Sandvik2016AC,shao2017SAC,Shao2023SAC} have been developed during the past decades, aiming at solving this problem~\cite{yan2022triangular,zhou2022quantum,yan2021topological,zhou2023quantum,yan2023unlocking,zhou2021amplitude,liu2022bulk}. Unfortunately, massive computing resources are required to get excitation spectrum. Moreover, these algorithms need to fit each spectrum case by case, with modifications that may lead to ambiguous results, not to mention the fitting process itself could be time-consuming. As a result, the computation complexity of numerical, analytical continuation methods limits these algorithms' power to reach larger lattice sizes, explore vast choices of parameters, or test various candidate materials. Is introducing a faster approach with less cost to extract energy-momentum dispersion from quantum Monte Carlo simulations feasible? In this paper, we show a possible access to large-scale calculation of the energy dispersion: the single-mode approximation (SMA) that has been widely used in the field of Bose-condensed systems, quantum information, quantum spin systems and condensed-matter theory~\cite{feynman1954atomic,griffin2009bose,toennies2004superfluid,haegeman2013elementary,yi2002single,bruschi2010unruh,Lauchli2008SAC,yan2022height}.

As far as we know, SMA has yet to be used in the QMC simulations. In this paper, we develop an efficient scheme extending the SMA into QMC algorithms to extract the dispersion information straightforwardly with extremely cheap computational cost and low barrier of technique. This new approach can reach large spin systems with up to $10^6$ spins and detect information of excitation in strongly correlated systems hosting phase transitions.

This paper is organized as follows: We begin in Sec. II by introducing the SMA. In Sec. III, we describe how the SSE algorithm cooperates with SMA, and how SMA works with the projector QMC method on a valence-bond (VB) basis. We show the results of several calculations in Sec. IV and conclude with a summary in Sec. V.

\section{Single-Mode Approximation}
SMA was first introduced by Richard Feynman in 1954 to investigate excited states in liquid helium. He estimated the lowest collective excitation energy of superfluid $^4$He by using this approach~\cite{feynman1954atomic}. This method has been widely applied to various systems including not only liquid helium but also for cold atom 
systems, 
BCS-BEC crossover, phonons in crystals, metals, and quantum spin systems~\cite{griffin2009bose,toennies2004superfluid,haegeman2013elementary,manousakis1991spin,yi2002single,bruschi2010unruh,Lauchli2008SAC,chen2005bcs,takeno1972theory}. In this paper, we mainly talk about how this method can be applied to spin-lattice models.

The key point of SMA is the assumption that by acting some momentum-dependent operators on the ground state, a single excitation can be created. An appropriate trial operator produces a well-estimated upper bound of the low-energy excitation~\cite{feynman1954atomic} in the long wave-length limit.

Naturally, in our lattice spin system, we choose the $S^z$ operator in momentum space as the form of excitation:
\begin{equation}
    \hat{S}^z(\boldsymbol{q}) = \frac{1}{\sqrt{N}} \sum_{i} e^{-i\boldsymbol{q} \cdot \boldsymbol{r}_i} \hat{S}^z_i,
    \label{eq:FourierTransformSz}
\end{equation}
where $\boldsymbol{q}$ is a given momentum. $\hat{S}^z_i$ denotes the $z$ component of the spin at site $i$. The approximated wavefunction to describe a lowest excited state is 
\begin{equation}
    \vert \psi_{\boldsymbol{q}} \rangle = \hat{S}^z(\boldsymbol{q}) \vert \mathrm{GS} \rangle.
    \label{eq:excitedState}
\end{equation}
If this state is orthogonal to the ground state, then the corresponding norm of this wavefunction is
\begin{equation}
    S_z^2(\boldsymbol{q}) = \langle \psi_{\boldsymbol{q}} \vert \psi_{\boldsymbol{q}} \rangle = \langle \mathrm{GS} \vert \hat{S}^{z\dagger}(\boldsymbol{q}) \hat{S}^z(\boldsymbol{q}) \vert \mathrm{GS} \rangle.
    \label{eq:norm}
\end{equation}

By using these notations, the SMA dispersion can be expressed as
\begin{equation}
    \omega_{\mathrm{SMA}} = \frac{\langle \psi_{\boldsymbol{q}} \vert (\hat{H} - E_0) \vert \psi_{\boldsymbol{q}} \rangle}{\langle \psi_{\boldsymbol{q}} \vert \psi_{\boldsymbol{q}} \rangle},
    \label{eq:variational}
\end{equation}
where $\omega_{\mathrm{SMA}}$ is an upper bound of energy gap at wave vector $\boldsymbol{q}$ and $E_0$ is the exact ground state energy~\cite{feynman1954atomic}.
Using Eq. (\ref{eq:excitedState}) and (\ref{eq:norm}), we can rewrite Eq.(\ref{eq:variational}) as
\begin{equation}
    \begin{aligned}
        \omega_{\mathrm{SMA}} & =  \frac{1}{S_z^2(\boldsymbol{q})} \langle \psi_{\boldsymbol{q}} \vert (\hat{H} - E_0) \vert \psi_{\boldsymbol{q}} \rangle \\& = \frac{1}{S_z^2(\boldsymbol{q})} \langle \mathrm{GS} \vert \hat{S}^{z}(-\boldsymbol{q}) [\hat{H}, \hat{S}^z(\boldsymbol{q})] \vert \mathrm{GS} \rangle \\& = \frac{1}{S_z^2(\boldsymbol{q})} \langle \mathrm{GS} \vert [\hat{S}^z(\boldsymbol{q}), \hat{H}] \hat{S}^z(\boldsymbol{-q}) \vert \mathrm{GS} \rangle \\& = \frac{1}{2} \frac{\langle \mathrm{GS} \vert [\hat{S}^{z}(-\boldsymbol{q}), [\hat{H}, \hat{S}^z(\boldsymbol{q})]] \vert \mathrm{GS} \rangle}{S_z^2(\boldsymbol{q})},
        \label{eq:SMA}
    \end{aligned}
\end{equation}
which completes our derivation of excitation dispersion of SMA.

In cases when state $|\Psi_{\boldsymbol{q}} \rangle $ is not orthogonal to the ground state $\vert \mathrm{GS} \rangle$, namely 
\begin{equation}
    \langle \mathrm{GS} \vert \psi_{\boldsymbol{q}} \rangle = \langle \mathrm{GS} \vert \hat{S}^z(\boldsymbol{q}) \vert \mathrm{GS} \rangle = c_1 \neq 0,
\end{equation}
we can express our wave function as
\begin{equation}
    \vert \psi_{\boldsymbol{q}} \rangle = c_1 \vert \mathrm{GS} \rangle + c_2 \vert \mathrm{ES} \rangle
\end{equation}
where $\vert \mathrm{ES} \rangle$ represents an excited-state orthogonal to the ground state
\begin{equation}
    \langle \mathrm{GS} \vert \mathrm{ES} \rangle = 0.
\end{equation}
To get the correct estimate of dispersion in such cases, one has to modify the approximation equation (\ref{eq:variational}) to
\begin{equation}
    \omega_{\mathrm{SMA}} = \frac{\langle \mathrm{ES} \vert (\hat{H} - E_0) \vert \mathrm{ES} \rangle}{\langle \mathrm{ES} \vert \mathrm{ES} \rangle}.
    \label{eq:variationalModified}
\end{equation}
After making use of the relation
\begin{equation}
    c_2^*c_2 \langle \mathrm{ES} \vert (\hat{H} - E_0) \vert \mathrm{ES} \rangle = \langle \psi_{\boldsymbol{q}} \vert (\hat{H} - E_0) \vert \psi_{\boldsymbol{q}} \rangle
\end{equation}
and the Gram-Schmidt process
\begin{equation}
    c_2 \vert \mathrm{ES} \rangle = \vert \psi_{\boldsymbol{q}} \rangle - \frac{\langle \mathrm{GS} \vert \psi_{\boldsymbol{q}} \rangle}{\langle \mathrm{GS} \vert \mathrm{GS}\rangle} \vert \mathrm{GS} \rangle,
\end{equation}
we express the final SMA expression as
\begin{equation}
    \begin{aligned}
        \omega_{\mathrm{SMA}} & = \frac{\langle \psi_{\boldsymbol{q}} \vert (\hat{H} - E_0) \vert \psi_{\boldsymbol{q}} \rangle}{(\langle \psi_{\boldsymbol{q}} \vert - \langle \mathrm{GS} \vert \frac{\langle \mathrm{GS} \vert \psi_{\boldsymbol{q}} \rangle}{\langle \mathrm{GS} \vert \mathrm{GS}\rangle})(\vert \psi_{\boldsymbol{q}} \rangle - \frac{\langle \mathrm{GS} \vert \psi_{\boldsymbol{q}} \rangle}{\langle \mathrm{GS} \vert \mathrm{GS}\rangle} \vert \mathrm{GS} \rangle)} \\& = \frac{1}{2} \frac{\langle \mathrm{GS} \vert [\hat{S}^z(-\boldsymbol{q}),[\hat{H}, \hat{S}^z(\boldsymbol{q})]] \vert \mathrm{GS} \rangle}{S_z^2(\boldsymbol{q}) - \frac{\vert \langle \mathrm{GS} \vert \psi_{\boldsymbol{q}} \rangle \vert^2}{\langle \mathrm{GS} \vert \mathrm{GS} \rangle}}.
        \label{eq:SMAModified}
    \end{aligned}
\end{equation}

In summary, the spirit of SMA is to construct a low-energy-excitation state, e.g., a spin-wave perturbated wave function as Eq. (\ref{eq:excitedState}), which is orthogonal to the ground state to estimate the upper bound of the first excited gap, thus $\omega_{\mathrm{SMA}} \geq \omega$. The ``$=$" holds only if the excited mode is single, then we have $\omega_{\mathrm{SMA}} = \omega$.

\begin{figure*}[!t]
    \centering
    \includegraphics[width = 0.75\textwidth]{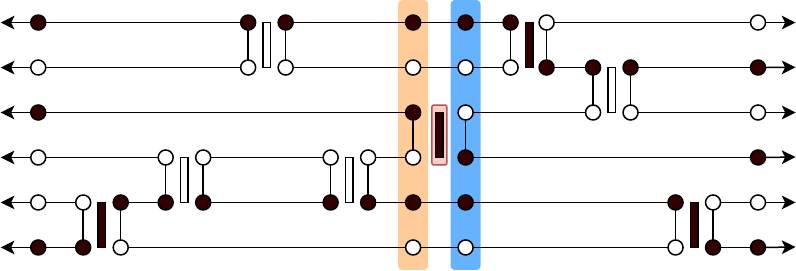}
    \caption{Sketch of SMA excitation dispersion measurement of a six-spin system with SSE. The imaginary time dimension is depicted horizontally. Filled and open circles represent $\up$ and $\down$ spins, respectively. Black bonds, which change the spin configuration, denote off-diagonal operators. White bonds that keep the spins unchanged denote diagonal operators. The identity operator is omitted in this figure. The bond highlighted by the red box represents the randomly chosen nonidentity operator. ``Configuration 1" and ``Configuration 2" denote the spin configurations in light orange and blue boxes, respectively. Here, ``Configuration 1" is $\vert \up \down \up \down \up \down \rangle$ and ``Configuration 2" is $\vert \up \down \down \up \up \down \rangle$.}
    \label{fig:imaginarytime}
\end{figure*}

SMA extracts the excitation information of the constructed operators, e.g., $\hat{S}^z(\boldsymbol{q})$. It provides a well approximated result where the spectrum displays little continuum, i.e., the spectrum is sharp. When there exists board continuum, the SMA quantity constitutes an upper bound of the continuum spectrum. The Ref.~\cite{Lauchli2008SAC} shows several examples to support this conclusion via comparing the NAC results with SMA results, in which the SMA and NAC use the same operators for the excitation. In conclusion, SMA can well describe the sharp excitation where one single mode of excitation nearly exhaust the full spectrum. When the excitation is not narrow, SMA gives its upper bound. 

Although it is believed that the SMA fails to give physical results when applied to strongly correlated systems, we show in Sec. IV that SMA is able to capture correctly the change of excitation gap when a phase transition happens. This fact indicates that SMA is still a useful tool to study exotic phases and phase transitions among them in strongly correlated systems.

\section{SMA combined with Quantum Monte Carlo}
\subsection{SMA combined with SSE}
In this section we introduce how to perform SMA calculations with SSE and explains how the measurements are performed.

The stochastic series expansion ~\cite{Sandvik1991SSE,Sandvik1999SSE,Syljuasen2002DirectedLoop,Sandvik2019SSE,Nisheeta2021SSE} approach constitutes a method to simulate sign-problem-free spin systems using quantum Monte Carlo techniques. Here, we briefly summarize the important part of this algorithm. Its starting point is the partition function of the system:
\begin{equation}
    Z = \mathrm{Tr} (e^{-\beta \hat{H}}).
\end{equation}
Its Taylor expansion replaces the exponential operator in the partition function, and the trace is rewritten as a summation over a complete basis of the system,
\begin{equation}
    Z = \sum_{\alpha} \sum_{n=0}^{\infty} \frac{\beta^n}{n!} \langle \alpha \vert (-\hat{H})^n \vert \alpha \rangle
\end{equation}
The Hamiltonian is written as the sum of several terms
\begin{equation}
    \hat{H} = \sum_{i} \hat{H}_i
\end{equation}
where $i$ is a label for enumerating different terms. The Taylor series is truncated at $M$. $M$ should be sufficiently large so that the truncation error is small enough and negligible. After all these steps, the partition function is
\begin{equation}
    Z = \sum_{\alpha} \sum_{\{ H_{a} \}} \frac{\beta^n (M-n)!}{M!} \langle \alpha \vert \prod_{i}\hat{H}_{a_i} \vert \alpha \rangle
\end{equation}
All possible operator strings with lengths between 0 and $M$ are summed over. $\alpha$ and $H_a$ are sampled during a Monte Carlo procedure according to each term's weight.

Since the SSE is usually performed in the spin $S_z$ basis, the equal-time correlation function of the spin $z$ component in Eq.(\ref{eq:norm}), i.e., the denominator part of Eq.(\ref{eq:SMA}), can be directly measured. The double commutator, i.e., the numerator part of Eq.(\ref{eq:SMA}), can be measured as follows. It contains four terms,
\begin{equation}
    \frac{1}{2} \langle \hat{S}_{\boldsymbol{q}}^{z\dag} \hat{H} \hat{S}_{\boldsymbol{q}}^{z} - \hat{S}_{\boldsymbol{q}}^{z\dag} \hat{S}_{\boldsymbol{q}}^{z} \hat{H} - \hat{H} \hat{S}_{\boldsymbol{q}}^{z} \hat{S}_{\boldsymbol{q}}^{z\dag} + \hat{S}_{\boldsymbol{q}}^{z} \hat{H} \hat{S}_{\boldsymbol{q}}^{z\dag} \rangle.
\end{equation}

We now briefly describe how to measure these quantities from SMA. After the system has reached equilibrium (ground state in this case), randomly choose a nonidentity operator from imaginary time. For convenience, we assume the imaginary time dimension is horizontal, as shown in Fig.\ref{fig:imaginarytime}. ``Configuration 1/2" (notated as ``C1/C2") denotes the state (or spin configuration) on the left/right side of the chosen operator, respectively. Notice that $\hat{S}_{\boldsymbol{q}}^{z}$ is diagonal (although not Hermitian in most cases) in the $S_z$ basis, so these $\hat{S}_{\boldsymbol{q}}^{z}$ and $\hat{S}_{\boldsymbol{q}}^{z\dag}$ operators applied on ``C1" or ``C2" result in complex numbers $S_{\boldsymbol{q}}^{z*}(\mathrm{C1})$ and $S_{\boldsymbol{q}}^{z}(\mathrm{C2})$:
\begin{equation}
    \begin{aligned}
        \langle \hat{S}_{\boldsymbol{q}}^{z\dag} \hat{H} \hat{S}_{\boldsymbol{q}}^{z} \rangle & = \langle S_{\boldsymbol{q}}^{z*}(\mathrm{C1}) S_{\boldsymbol{q}}^{z}(\mathrm{C2}) \hat{H} \rangle \\& =\langle S_{\boldsymbol{q}}^{z*}(\mathrm{C1}) S_{\boldsymbol{q}}^{z}(\mathrm{C2}) \frac{\hat{n}}{\beta} \rangle
        \label{eq:SMA_SSE}
        \end{aligned}
\end{equation}
Notice that we use a ``hat" notation to distinguish quantum operators from numbers. In the second line, $\hat{H}$ is replaced by $\hat{n} / \beta$, which is the energy estimator in the SSE algorithm~\cite{Sandvik1999SSE,Sandvik2019SSE}. Making use of the relation Eq. (\ref{eq:SMA_SSE}), the double commutator estimator can be expressed as
\begin{equation}
   \begin{aligned}
       f(\boldsymbol{q}) = \frac{1}{2} \langle  (& S_{\boldsymbol{q}}^{z*}(\mathrm{C1}) S_{\boldsymbol{q}}^{z}(\mathrm{C2}) - S_{\boldsymbol{q}}^{z*}(\mathrm{C1}) S_{\boldsymbol{q}}^{z}(\mathrm{C1}) \\& - S_{\boldsymbol{q}}^{z}(\mathrm{C2}) S_{\boldsymbol{q}}^{z*}(\mathrm{C2}) + S_{\boldsymbol{q}}^{z}(\mathrm{C1}) S_{\boldsymbol{q}}^{z*}(\mathrm{C2}))\times\frac{\hat{n}}{\beta} \rangle,
   \end{aligned}
\end{equation}
which is the final expression of the SMA dispersion. This expression is independent of forms of Hamiltonians, enabling us to use SMA in systems with complicated Hamiltonians.

\begin{figure}[!t]
    \centering
    \includegraphics[width = 0.3\textwidth]{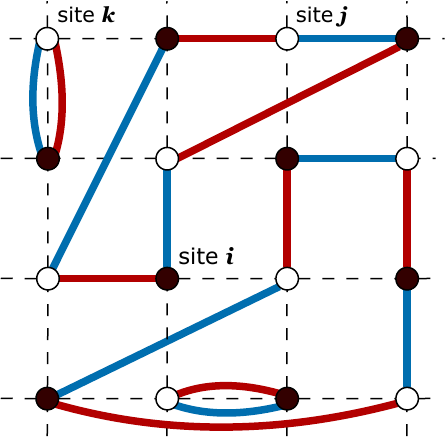}
    \caption{Illustration of a valence-bond transposition graph on a $4 \times 4$ square lattice. This valence bond basis is used in projector QMC approach. Solid circles represent sites on sublattice A, and open circles represent sites on sublattice B. Red and blue bond configuration represent the bra $\langle \psi_{\mathrm{L}} \vert$ and the ket $\vert \psi_{\mathrm{R}} \rangle$, respectively. This figure is a transposition graph of the inner product $\langle \psi_{\mathrm{L}} \vert \psi_{\mathrm{R}} \rangle$. Spins on site $i$ and $j$ belong to the same loop, so $\langle \psi_{\mathrm{L}} \vert \hat{\boldsymbol{S}}_i \cdot \hat{\boldsymbol{S}}_j \vert \psi_{\mathrm{R}} \rangle / \langle \psi_{\mathrm{L}} \vert \psi_{\mathrm{R}} \rangle = - 3/4$. Spins on-site $i$ and $k$ are not correlated since they are in different loops.}
    \label{fig:valencebond}
\end{figure}

\subsection{SMA combined with projector QMC}
In cases where the spin system only involves Heisenberg interaction that preserve SU(2) spin symmetry, we can combine SMA with projector QMC. Resulted method can be easily parallelized, making it faster and more efficient. 
The Projector QMC method was initially introduced by Sandvik~\cite{Sandvik2005Projector,Sandvik2010Projector} to access ground states of quantum spin systems efficiently. This algorithm is formulated in a combined space of spin $S_z$ and valence-bond bases. Since one can directly obtain information on valence bonds, it is naturally suited for studying spin rotationally invariant Hamiltonians, such as the Heisenberg model and its extension versions with long-range interactions. The properties of valence-bond basis and projector QMC methods have been demonstrated in detail in the literature~\cite{Sandvik2005Projector,Sandvik2010Projector,Beach2006VBbasis}.

In this section, we describe our algorithm with the Heisenberg model
\begin{equation}
    \hat{H} = \sum_{i,j} J_{ij} \hat{\boldsymbol{S}}_i \cdot \hat{\boldsymbol{S}}_j.
    \label{eq:heisenberg}
\end{equation}

Based on the definition of SMA dispersion, we define
\begin{equation}
    f(\boldsymbol{q}) = \frac{1}{2} \langle \mathrm{GS} \vert [\hat{\boldsymbol{S}}(-\boldsymbol{q}), [\hat{H}, \hat{\boldsymbol{S}}(\boldsymbol{q})]] \vert \mathrm{GS} \rangle, \label{eq:numerator}
\end{equation}
where $\hat{\boldsymbol{S}}(\boldsymbol{q})$ denotes the Fourier transform of spin operator
\begin{equation}
    \hat{\boldsymbol{S}}(\boldsymbol{q}) = \frac{1}{\sqrt{N}} \sum_{i} e^{-i \boldsymbol{q} \cdot \boldsymbol{r}_i} \hat{\boldsymbol{S}}_i.
    \label{eq:FourierTransformS}
\end{equation}
After we expanding Eq.(\ref{eq:numerator}) with Eq.(\ref{eq:FourierTransformS}) and Hamiltonian Eq.(\ref{eq:heisenberg}), $f(\boldsymbol{q})$ is written as
\begin{equation}
    \begin{aligned}
        f(\boldsymbol{q}) = \frac{1}{2N} & \sum_{i,j} J_{ij} \sum_{l,l'} e^{-i \boldsymbol{q} \cdot (\boldsymbol{r}_l - \boldsymbol{r}_{l'})} \\& \langle \mathrm{GS} \vert [\hat{\boldsymbol{S}}_{l'}, [\hat{\boldsymbol{S}}_i \cdot \hat{\boldsymbol{S}}_j, \hat{\boldsymbol{S}}_l]] \vert \mathrm{GS} \rangle.
        \label{eq:numeratorExpansion}
    \end{aligned}
\end{equation}
We note that all multiplications in the commutation relation are dot products of vectors. Only terms of which both subscripts $l$ and $l'$ take values $i$ or $j$ are nonzero since operators on different sites commute with each other. Commutation relations~\cite{Voros2021SUN}
\begin{equation}
    [\hat{\boldsymbol{S}}_i, [\hat{\boldsymbol{S}}_i \cdot \hat{\boldsymbol{S}}_j, \hat{\boldsymbol{S}}_i]] = - 2 \hat{\boldsymbol{S}}_i \cdot \hat{\boldsymbol{S}}_j,
\end{equation}
\begin{equation}
    [\hat{\boldsymbol{S}}_i, [\hat{\boldsymbol{S}}_i \cdot \hat{\boldsymbol{S}}_j, \hat{\boldsymbol{S}}_j]] = 2 \hat{\boldsymbol{S}}_i \cdot \hat{\boldsymbol{S}}_j
\end{equation}
can be obtained after some simple $\mathrm{SU}(2)$ algebra calculations. Taking all of these relations into account, we finally get
\begin{equation}
    \begin{aligned}
        f(\boldsymbol{q}) = - \frac{4}{N} \sum_{i,j} J_{ij} \sin^2(\frac{1}{2} \boldsymbol{q} \cdot (\boldsymbol{r}_i - \boldsymbol{r}_j)) \langle \hat{\boldsymbol{S}}_i \cdot \hat{\boldsymbol{S}}_j \rangle_{\mathrm{GS}}.
    \end{aligned}
\end{equation}
Here we have replaced $\langle \mathrm{GS} \vert \hat{\boldsymbol{S}}_i \cdot \hat{\boldsymbol{S}}_j \vert \mathrm{GS} \rangle$ by its abbreviation $\langle \hat{\boldsymbol{S}}_i \cdot \hat{\boldsymbol{S}}_j \rangle_{\mathrm{GS}}$. Spin-spin correlation in the ground state can be easily estimated in the VB basis~\cite{Beach2006VBbasis}:
\begin{equation}
    \langle \hat{\boldsymbol{S}}_i \cdot \hat{\boldsymbol{S}}_j \rangle_{\mathrm{GS}} = \frac{3}{4} \epsilon_{ij} \delta^{ij}.
    \label{con:correlatorVB}
\end{equation}
Here $\epsilon_{ij}$ is one if site $i$ and $j$ are on the same sublattice of underlying bipartite lattice, and is $-1$ otherwise. $\delta^{ij}$ equals one if site $i$ and $j$ belong to the same loop formed in the transposition graph. If they belong to different loops, then $\delta^{ij}$ takes the value of 0. See Fig.\ref{fig:valencebond} as an example.

To obtain the SMA dispersion of the Heisenberg model, only spin-spin correlations in the ground state are necessary. Projector QMC and VB basis offer quick and convenient access to these correlation functions. Non-trivial parallel programming can be applied to this algorithm, remarkably enhancing power and efficiency. These advantages enable us to obtain a dispersion of systems with $10^4$, even $10^5$ spins.

Although the dispersion of some simple Hamiltonians, like the Heisenberg model, can be simulated and obtained easily, the method discussed in this section is not suitable for systems with complicated Hamiltonians. Trying to simplify the double commutator may eventually obtain observables that are hard to estimate in practice. For example, $Q$ term interaction $(\hat{\boldsymbol{S}}_i \cdot \hat{\boldsymbol{S}}_j)(\hat{\boldsymbol{S}}_k \cdot \hat{\boldsymbol{S}}_l)$ in $J$-$Q$ model ~\cite{Sandvik2007ProjectorJQ,Sandvik2010JQ,Pujari2013ProjectorJQ} produce terms including
\begin{equation}
    [\hat{\boldsymbol{S}}_i, [(\hat{\boldsymbol{S}}_i \cdot \hat{\boldsymbol{S}}_j)(\hat{\boldsymbol{S}}_k \cdot \hat{\boldsymbol{S}}_l), \hat{\boldsymbol{S}}_k]] = (\hat{\boldsymbol{S}}_i \times \hat{\boldsymbol{S}}_j) \cdot (\hat{\boldsymbol{S}}_k \times \hat{\boldsymbol{S}}_l).
\end{equation}
Cross-product terms can be estimated using QMC. However, the procedure would be cumbersome since cross-product terms contain several off-diagonal operators. Thus, projector QMC with SMA does not fit when $J$-$Q$ model is of interest. 

We conclude that the principle is, if the double commutator can be simplified into an easily calculated estimator in valence-bond basis or $S_z$ basis, then projector QMC with SMA is able to solve this problem very efficiently, since projector QMC can be performed with nontrivial programming. If ideal simplification cannot be achieved, then SSE with SMA should be applied, since SSE + SMA is more general.

The following section will show several examples calculated using the methods mentioned above.

\section{Results}
\subsection{Two-dimensional AFM Heisenberg Model}

\begin{figure}[!t]
    \centering
    \includegraphics[width = 0.35\textwidth]{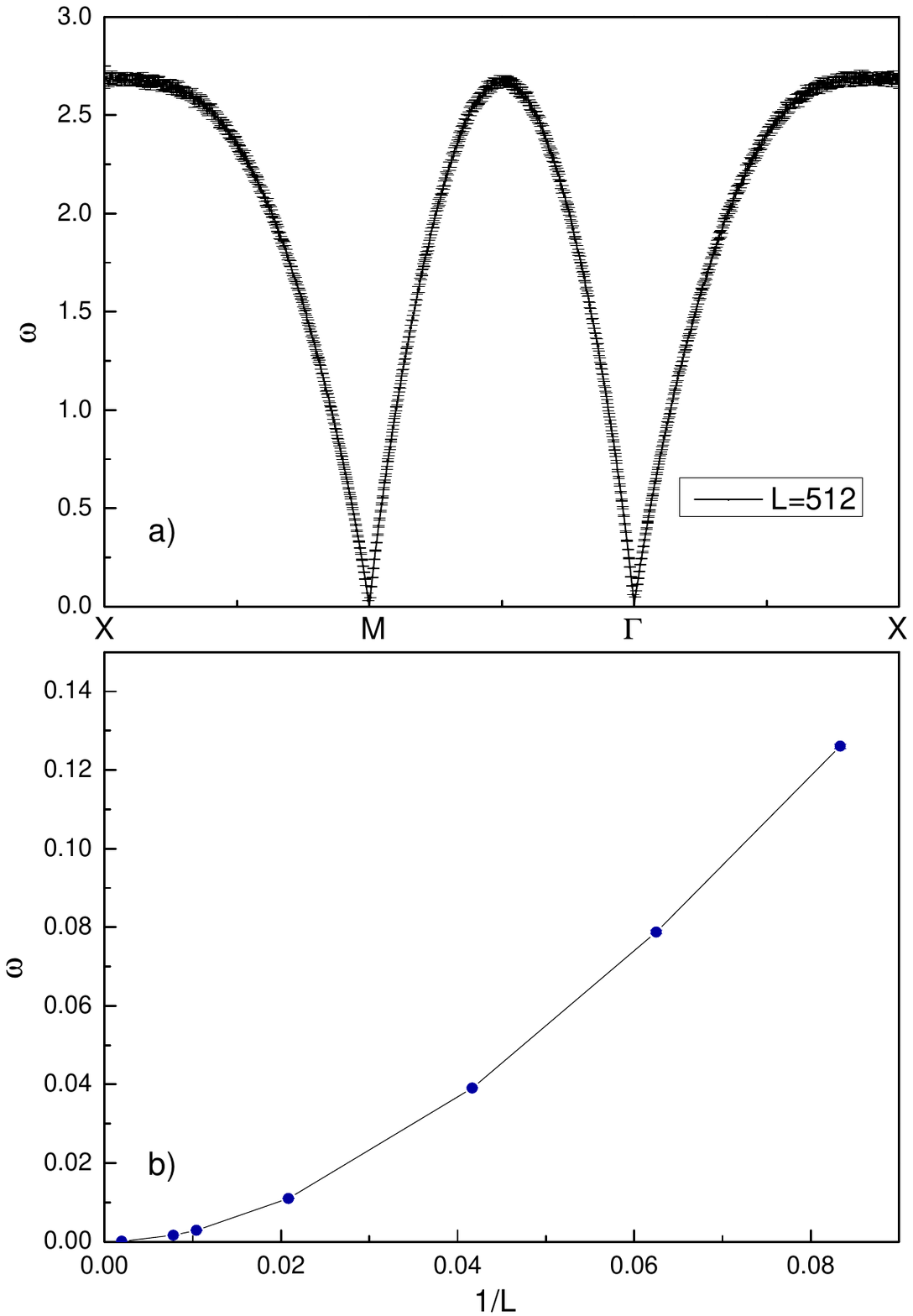}
    \caption{(a) SMA dispersion of 2D AFM Heisenberg model on a square lattice of system size $L = 512$. Periodic boundary condition is applied. This dispersion is obtained from projector QMC combined with SMA. Two gapless excitation modes exist at $M$ and $\Gamma$ points, respectively. (b) Energy excitation gap at $M$ point of different system sizes. With the increase in system size, the gap at $M$ converges to zero. Errors of data are smaller than the symbol sizes.}
    \label{fig:AFM}
\end{figure}

The first case is the two-dimensional (2D) antiferromagnetic (AFM) Heisenberg model with only nearest-neighbor interactions on a square lattice
\begin{equation}
    \hat{H} = J \sum_{\langle i, j \rangle} \hat{\boldsymbol{S}}_i \cdot \hat{\boldsymbol{S}}_j
\end{equation}
where $\langle i, j \rangle$ denotes nearest-neighbor sites, and coupling $J > 0$.

This Hamiltonian is simple, so we calculate using projector QMC with nontrivial parallel programming. In the simulation, we set the imaginary time length $m = 0.8 L^3$, length of measurement 100,000 times $\times$ 40 bins. Systems with $512 \times 512$ spins are simulated, and the SMA dispersion obtained is shown in Fig. \ref{fig:AFM}(a). Two gapless modes exist, one at $M$ point of momentum space and the other at $\Gamma$ point. Both gapless modes have linear dispersion in the low-energy part. This result is consistent with the dispersion given by spin-wave theory~\cite{dalla2015excitation,Song2023LongRange}.

It is worth noting here that indicated by the original SMA expression Eq.(\ref{eq:SMA}), either the double commutator vanishes or the equal-time correlation function diverges as a function of system size would induce the absence of an energy gap.

At the $\Gamma$ point, the operator acted on the system commutes with the total Hamiltonian,
\begin{equation}
    [\hat{S}_z(\boldsymbol{q}=0), J\sum_{\langle i, j \rangle} \hat{\boldsymbol{S}}_i \cdot \hat{\boldsymbol{S}}_j] = 0.
\end{equation}
As a result, the numerator in Eq.(\ref{eq:SMA}) is always zero, regardless of the system size.

At the $M$ point, the equal-time correlation function on the denominator increases with system size and finally diverges in the thermal-dynamic limit. This fact indicates that there must be a gapless mode at $M$. As shown in Fig.\ref{fig:AFM} (b), the energy gap becomes smaller and converges to zero with the increase of lattice size.

We mention here that the key point of numerical analytic continuation is fitting the dispersion according to the imaginary time correlation data. However, to get accurate information of the low-energy part of the dispersion using NAC method, one has to measure correlations of very long imaginary time distances with high precision, and fit the correlation data several times according to the value of the entropy or other standards. In fact, it needs high-technical barrier to write an extra code for the numerical analytic continuation in speciality. That is why we want to develop a method to obtain the dispersion quickly with low barrier.

The SAC method successfully obtained the dispersion function of 2D square-lattice AFM Heisenberg model~\cite{shao2017SAC}, which is also calculated here. The SAC can perform the continuum spectrum while SMA can only get a single dispersion. But the dispersion catches the main mode with the largest weight in the spectrum. Due to the low cost of SMA method, we can simulate much larger system size. Data of the dispersion of $512 \times 512$ 2D AFM model are provided, of which the size is far beyond the SAC method's reach (about $10^3$ spin systems~\cite{Shao2023SAC}).

\subsection{Two-dimensional long-range FM Heisenberg Model}
The next example is the 2D ferromagnetic (FM) Heisenberg model with long-range interactions. The Hamiltonian is
\begin{equation}
    \hat{H} = \sum_{i, j} J_{ij} \hat{\boldsymbol{S}}_i \cdot \hat{\boldsymbol{S}}_j
\end{equation}
with $J_{ij} < 0$. Here the term ``long-range" means that the coupling strength decays as a power-law form:
\begin{equation}
    \hat{H} = \sum_{i,j} \frac{1}{\vert r_{ij} \vert^{\alpha}} \hat{\boldsymbol{S}}_i \cdot \hat{\boldsymbol{S}}_j.
\end{equation}
The power exponent $\alpha$ controls the effective range of coupling. As $\alpha$ approaches infinity, the model returns to the Heisenberg model with only nearest-neighbor interactions. Strong long-distance couplings come in with small $\alpha$. Spectrums of ferromagnetic Heisenberg models can be well estimated by spin-wave theory~\cite{Song2023LongRange,Defenu2023LongRange,Diessel2023LongRange}. According to spin-wave theory, the dispersion of a magnon is
\begin{equation}
    \omega_{\mathrm{FM}}(\boldsymbol{q}) = \vert J_0 - J_{\boldsymbol{q}} \vert,
\end{equation}
where $J_{\boldsymbol{q}}$ is the Fourier transform of $J_{ij}$
\begin{equation}
    J_{\boldsymbol{q}} = \sum_{\boldsymbol{r}} e^{-i \boldsymbol{q} \cdot \boldsymbol{r}} J_{\boldsymbol{r}}.
\end{equation}

\begin{figure}[!t]
    \centering
    \includegraphics[width = 0.4\textwidth]{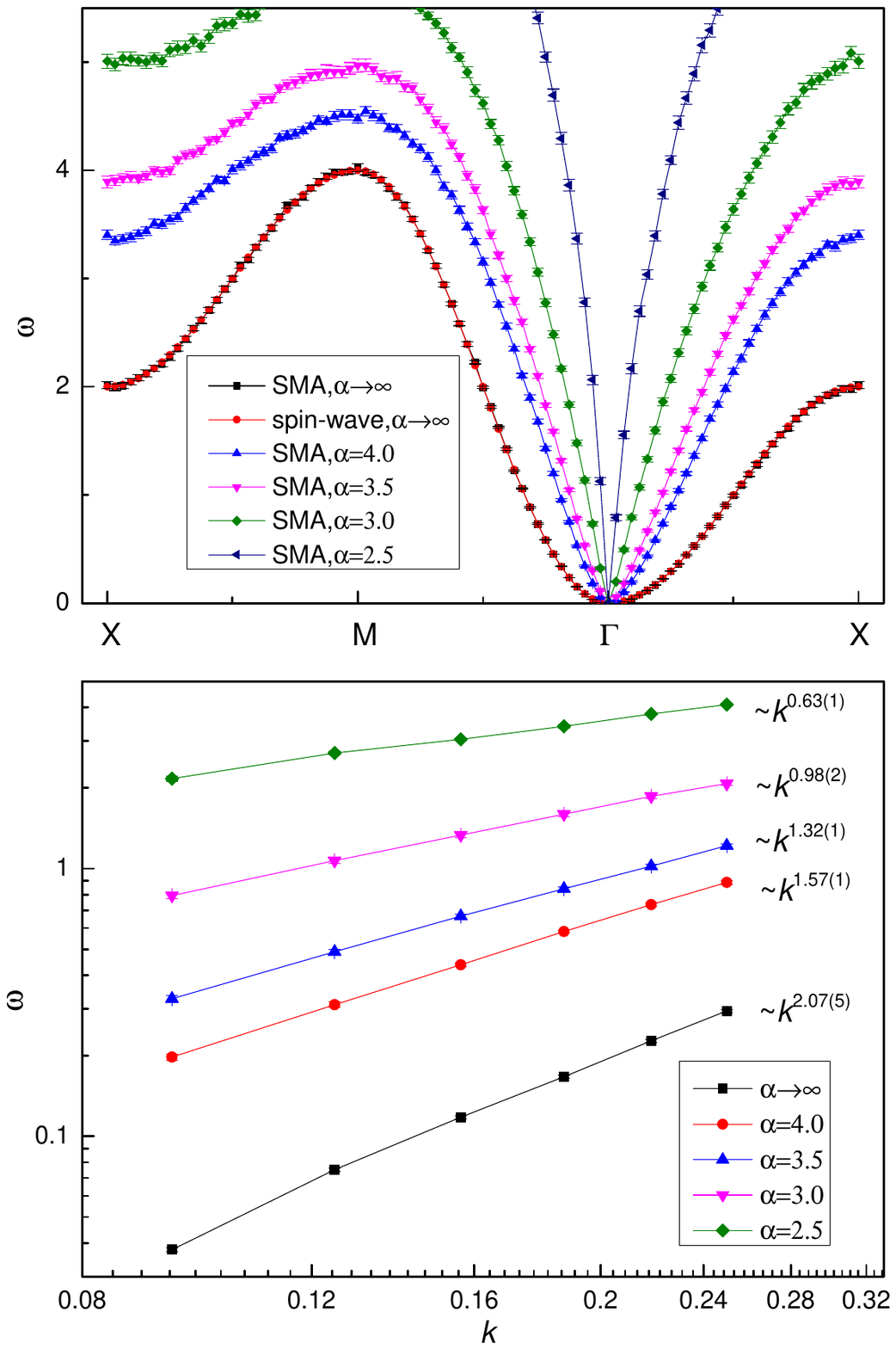}
    \caption{The upper panel is the SMA dispersion obtained by SSE simulation and spin-wave theory dispersion of the 2D ferromagnetic Heisenberg model on a square lattice. Lattice size $L = 48$. Periodic boundary condition is applied. Only interactions between nearest neighbors are included when power exponent $\alpha$ approaches infinity. The $\alpha \rightarrow \infty$ SMA results fit well with the spin-wave theory. Different $\alpha$ leads to different dispersion relation near $\Gamma$. The lower panel is the dispersion relations near the $\Gamma$ point. Power-law fitting results are labeled on the panel. The dispersion power exponent decreases with decreasing $\alpha$. Error bars are smaller than the size of the symbol.}
    \label{fig:FM}
\end{figure}

\begin{figure}[!t]
    \centering
    \includegraphics[width = 0.4\textwidth]{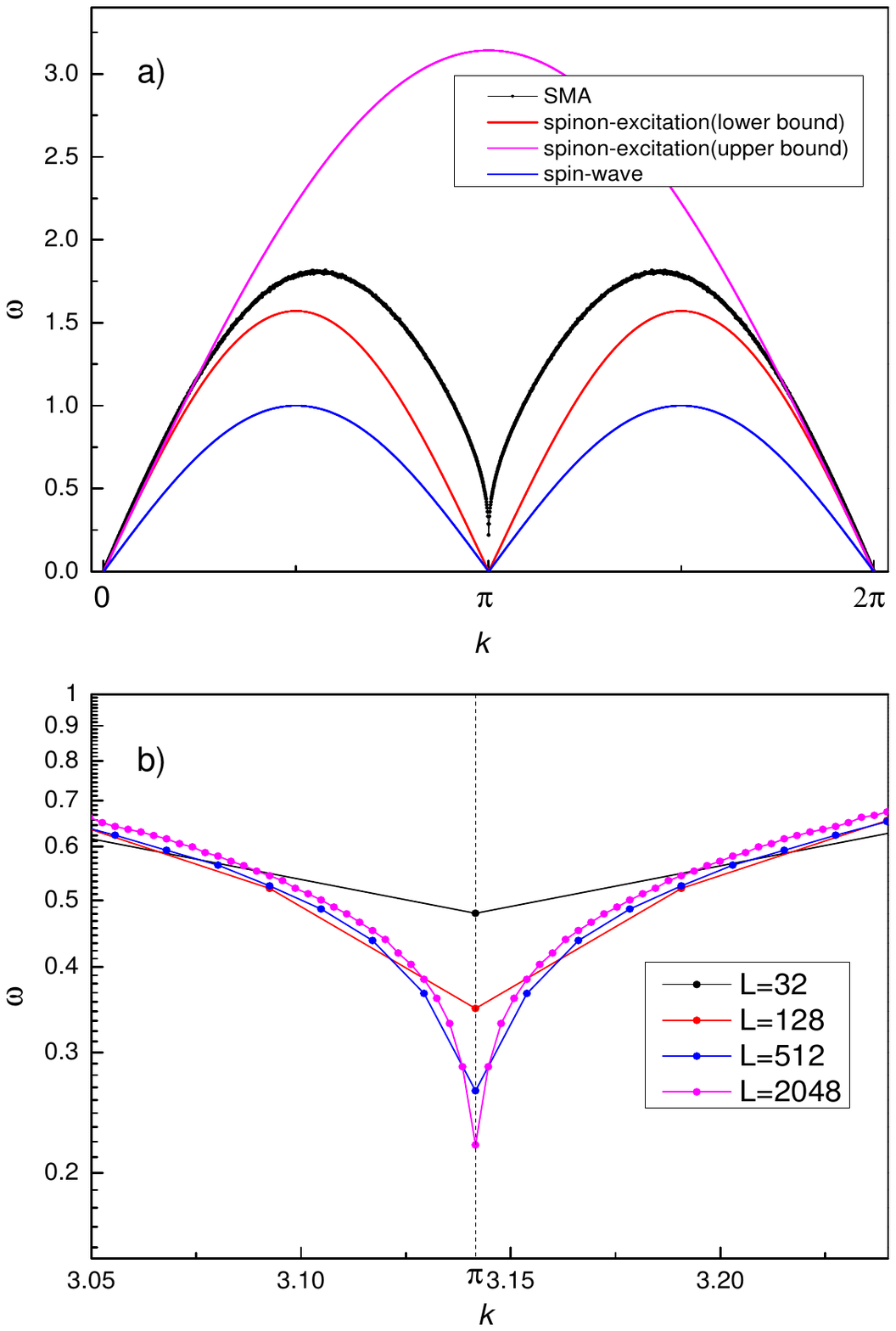}
    \caption{(a) The black line shows the SMA dispersion from SSE simulations of the antiferromagnetic Heisenberg chain. The lower and upper bounds of spinon excitation are shown with red and purple lines, respectively. The lower bound of spinon excitation is $\omega = \frac{1}{2} \pi \vert J \mathrm{sin} k\vert$, indicated by the red line. The upper bound of spinon excitation is $\omega = \pi \vert J \vert \mathrm{sin}\frac{1}{2} k$, which is shown by the purple line~\cite{takahashi1999thermodynamics}. The blue line indicates the result of the spin-wave theory (which is wrong). The length of the chain is $L = 2048$. (b) Energy gap near momentum $\pi$ of different chain length. Maximum chain length $L = 2048$ is reached. In both figures, error bars are smaller than the symbols.}
    \label{fig:chain}
\end{figure}

Here we compare our SSE with SMA results with the dispersion of magnon. We set $\beta = L$ here. Results are exhibited in Fig. \ref{fig:FM}. Figure \ref{fig:FM}(a) shows the dispersion of the Heisenberg model with different decay exponent $\alpha$. When $\alpha$ approaches infinity, only nearest-neighbor interactions are considered. In this case, our result is consistent with the spectrum given by spin-wave theory. The corresponding dispersion near the $\Gamma$ point is quadratic. As is shown in Fig. \ref{fig:FM}, as $\alpha$ decreases to $2.5$, this gapless mode still retains. However, the dispersion relations~\cite{Song2023LongRange}
\begin{equation}
    \omega \sim k^s
\end{equation}
varies with $\alpha$. In the nearest-neighbor version, dispersion power exponent $s = 2$. As $\alpha$ decreases, $s$ also decreases. This mode has a linear dispersion when $\alpha = 3.0$. Corresponding $s$ has been tagged on the lower panel of Fig. \ref{fig:FM}. All results are well compatible with magnon dispersion given by spin-wave theory for long-range interactions~\cite{Song2023LongRange,Diessel2023LongRange}.

\subsection{AFM Heisenberg Chain}
If the SMA algorithm always gives the same dispersion as spin-wave theory, undoubtedly, then it makes this method less appealing. Fortunately, this is not the case.

The next case shown in this paper is an AFM Heisenberg chain with periodic boundary conditions. The Hamiltonian is
\begin{equation}
    \hat{H} = J \sum_{i=1}^N \hat{\boldsymbol{S}}_i \cdot \hat{\boldsymbol{S}}_{i+1}
\end{equation}
where $\hat{\boldsymbol{S}}_{N+1} = \hat{\boldsymbol{S}}_1$ and $J > 0$.

As is known, spin-wave theory breaks down here and gives a wrong dispersion velocity~\cite{takahashi1999thermodynamics}
\begin{equation}
    v_{\mathrm{SW}} = \vert J \vert
\end{equation}
which is shown in Fig. \ref{fig:chain} with the blue line. This velocity is smaller than the correct result obtained from spinon theory~\cite{takahashi1999thermodynamics},
\begin{equation}
    v_{\mathbf{spinon}} = \frac{\pi}{2} \vert J \vert.
\end{equation}
In the simulation, we fix the temperature $\beta = 100$. As shown in Fig. \ref{fig:chain}, we can obtain the correct velocity near momentum $0$ and $2\pi$ from SSE with SMA calculations. In this case, SMA still works while spin-wave theory breaks down, indicating SMA calculation's better feasibility.

At momentum $\pi$, according to spinon excitation, there exists a strongly continuous spectrum~\cite{takahashi1999thermodynamics}. In such cases, SMA's upper-bound energy gap estimation is unreliable. With the increase of the system size, the gap given by SMA becomes smaller [Fig. \ref{fig:chain} (b)]. With the chain length increase, this gap converges to zero as the system approaches the thermodynamic limit. However, SMA does not tell the correct velocity of dispersion near momentum $\pi$ because of the continuous spectrum.

\begin{figure}[!t]
    \centering
    \includegraphics[width = 0.36\textwidth]{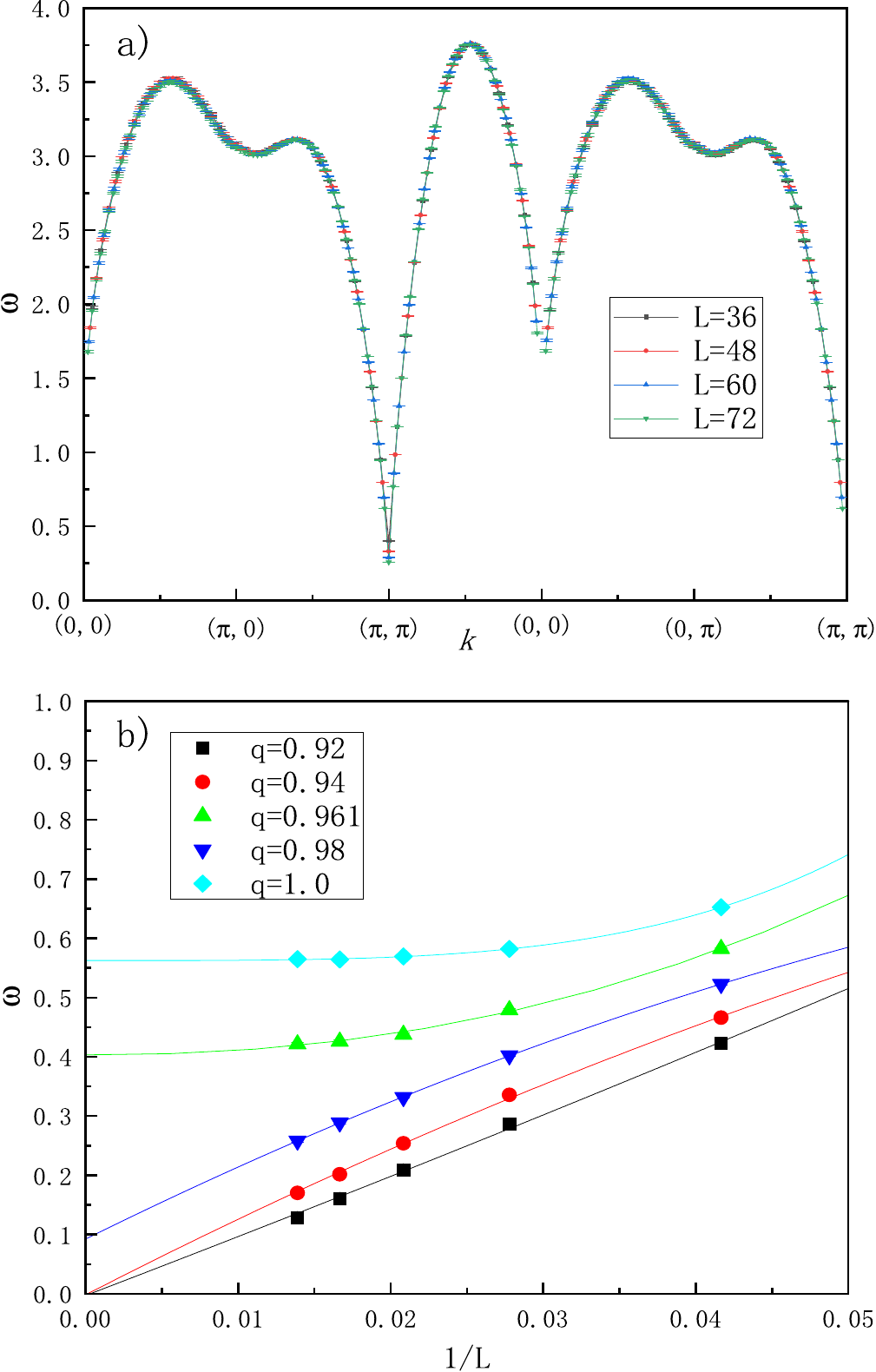}
    \caption{(a) SMA dispersion of $J$-$Q$ model at the phase transition point $q_c = 0.961$. Systems with sizes up to $72 \times 72$ are calculated. (b) Extrapolation of the dispersion value at $(\pi, \pi)$ point. When the system is in AFM phase ($q < 0.961$), the excitation is gapless. When the system enters cVBS phase ($q > 0.961$), there clearly exists a finite energy gap at $M$ point. At the phase transition point, the energy gap is not strictly zero because of the existence of continuum spectrum.}
    \label{fig:JQ}
\end{figure}

\subsection{Two-dimensional $J$-$Q$ Model}
In order to illustrate that the SMA is able to capture information of excitations in strongly correlated systems with competing states, we present our last model in this paper, the 2D $J$-$Q$ model~\cite{Sandvik2007ProjectorJQ,Sandvik2010JQ,takahashi2024so}. The Hamiltonian of this model is
\begin{equation}
    H = -J \sum_{\langle ij \rangle} \hat{P}_{ij} - Q \sum_{\langle \langle ijkl \rangle \rangle} \hat{P}_{ij} \hat{P}_{kl},
\end{equation}
where $\langle ij \rangle$ denotes nearest-neighbor sites and $\langle \langle ijkl \rangle \rangle$ denotes four corners of a plaquette. $ij$ and $kl$ are two parallel vertical or horizontal bonds of a plaquette. $\hat{P}_{ij}$ denotes the singlet projector operator on site $i$ and $j$,
\begin{equation}
    \hat{P}_{ij} = \frac{1}{4} - \hat{\boldsymbol{S}}_i \cdot \hat{\boldsymbol{S}}_j.
\end{equation}

$J$-$Q$ model hosts a weakly first-order phase transition between columnar valence bond solid (cVBS) state and N\'eel state~\cite{Sandvik2010JQ,deng2024diagnosing,takahashi2024so,JRZhao2021,song2024extracting}. The phase transition point is $q_c = [Q/(J+Q)]_c = 0.961$~\cite{lou2009antiferromagnetic}. The system has AFM order when $q < q_c$, and has cVBS order for $q > q_c$.

The low-energy excitations of AFM phase are gapless magnons at $\Gamma$ point (0, 0) and $M$ point ($\pi$, $\pi$), which we have already discussed in Sec. IV A, and shown in Fig. \ref{fig:AFM}. When $q$ approaches $q_c$, the excitations in the system gradually become spinons, which are $\frac{1}{2}$ fractionalized gapless excitations with continuum spectrum. This is a result of deconfined quantum criticality~\cite{senthil2004deconfined,takahashi2024so}. When the system crosses the phase transition point and turns into cVBS phase, the low-energy excitation becomes gapped triplons.

SMA dispersion of $J$-$Q$ model at the phase transition point is calculated with system sizes up to $72 \times 72$, illustrated in Fig.\ref{fig:JQ} (a). Two valleys, at the $M$ point and $\Gamma$ point respectively, are found, corresponding to the two gapless excitation modes of AFM phase. We calculate the energy gap at $M$ with the parameters around the phase transition point, and extrapolate the result to the thermodynamic limit. Results are shown in Fig.\ref{fig:JQ} (b), when the system is in N\'eel state, the excitation at $M$ point is gapless. For $q > q_c$ and the system is in cVBS phase, the energy gap converges to a finite value, which means the excitation is gapped. The energy gap at the phase transition point does not strictly converge to zero because of the continuum spectrum caused by fractionalized excitations. This result is consistent with the analysis in the preceding paragraph, which shows that SMA is able to correctly capture the excitation information to some extent even in a strongly correlated system.

\section{Conclusions}
What we have done in this paper is to combine the SMA into QMC simulations. We are not introducing a brand new spectrum-calculating process. As a result, the dispersion of excitations can be estimated with few extra efforts. 
We have to emphasize again that although the NAC can extract the spectrum from the data of imaginary time correlation functions while the SMA only gives dispersion, the NAC itself has a lot of tricks that not easy to manipulate, and the code of NAC is not easy to duplicate. On the other hand, the requirement of the imaginary time correlation function data is also very high. The NAC needs dense data of imaginary time correlations which also greatly increases the amount of computation.

The dispersion now can be obtained during the QMC simulations in this way. No extra processing and fitting is necessary. The measurement takes at most the same effort as the measurement of some simple observables, such as magnetization $M^2$, with a time complexity of O($N$) where $N$ represents the lattice size. Thus, the measurement of SMA can be done with little computing cost. The NAC needs extra fitting process to get the spectrum from accurate imaginary-time correlation with extremely long imaginary time distance~\cite{Shao2023SAC}. Obtaining the correlation needed with high precision is indeed time-consuming, which is actually the main reason preventing NAC to access larger system sizes.

The method introduced here has low technique barriers and is suitable for some large-scale simulation like scanning through parameter space, of which inverse methods are not capable. Actually when scanning through a parameter space, we do not need our results so accurate. One can use SMA to find some parameters where the behavior of the system may be interesting and then use NAC to calculate the dynamical spectrum in the vicinity of these parameters. 

We conclude here that we introduce an algorithm to perform SMA calculations via quantum Monte Carlo. In particular, two versions of the combination of SMA with quantum Monte Carlo are employed. Projector QMC with nontrivial parallel programming can be applied when directly simplifying the double commutator. In this case, large systems with $512\times 512$ spins are accessible in few days. For a system with a complicated Hamiltonian, we develop another general method with which forms of the Hamiltonians become irrelevant. Both algorithms can perform large-scale simulations outside of the reach of conventional spectrum-estimating algorithms. They may play an important role when large system sizes are crucial in exhibiting exotic excitations, and many systems should be selected according to their excitations. Several cases are calculated as examples. In the 2D Heisenberg model, either ferromagnetic or antiferromagnetic, SMA calculations give the correct excitation dispersions consistent with spin-wave theory. In the 1D antiferromagnetic chain, SMA goes beyond the spin-wave theory. Although the approximation near continuous spectrum could be more accurate, the correct velocity of dispersion near momentum 0 and $2\pi$ can be obtained. Furthermore, in the 2D $J$-$Q$ model, it is shown that SMA still works even when there exist strongly competing states. With the advent of this practical algorithm, scanning through parameters and performing statistical works of the dispersion have become possible.

\section{Acknowledgments}
This work is supported by the National Key Research and Development Program of China Grant No. 2022YFA1404204, and the National Natural Science Foundation of China Grant No. 12274086. Z.Y. thanks the inspirational discussions with Zi Yang Meng, Amos Chan, and David Huse in another related project and the support from the start-up funding of Westlake University and the open fund of Lanzhou Center for Theoretical Physics (12247101). Y.C.W. acknowledges the support from Zhejiang Provincial Natural Science Foundation of China (Grant No. LZ23A040003), and the support from the High-Performance Computing Centre of Hangzhou International Innovation Institute of Beihang University. The authors thank the high-performance computing center of Westlake University and the Beijng PARATERA Tech Co.,Ltd. for providing HPC resources.

\bibliography{SMA_QMC}

\end{document}